# Turing pattern theory on homogeneous and heterogeneous higher-order temporal network system FREE

Junyuan Shi; Linhe Zhu

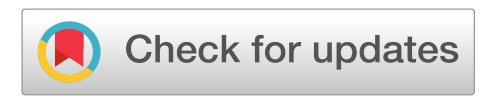

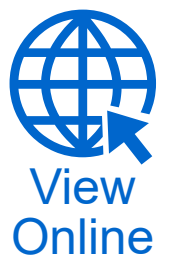

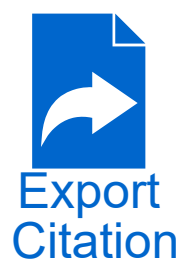

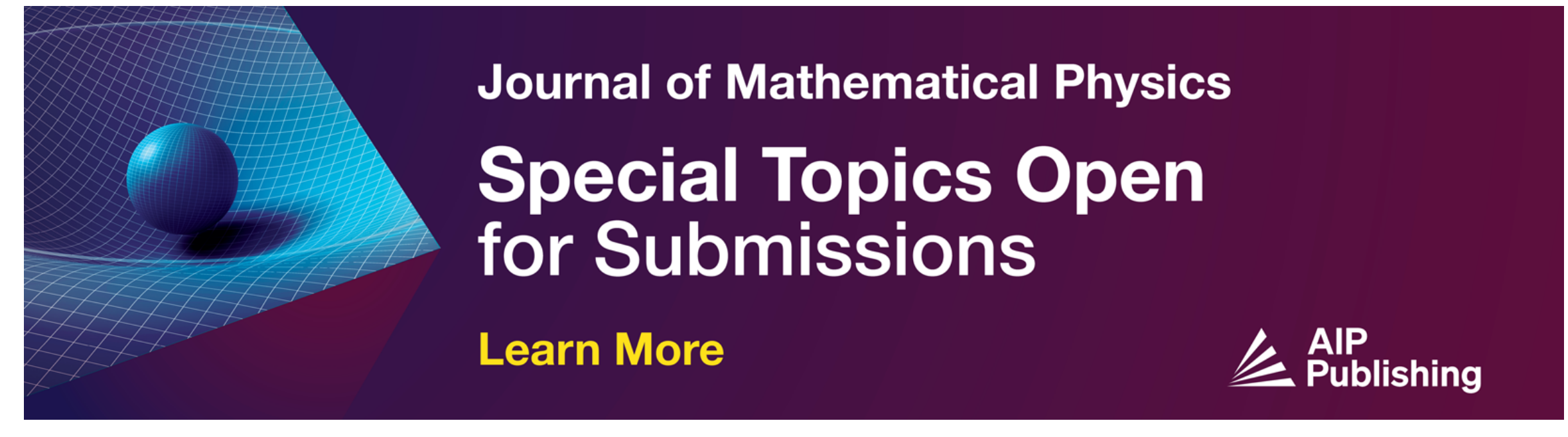

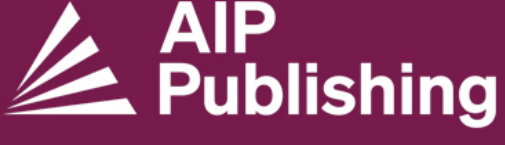

# Turing pattern theory on homogeneous and heterogeneous higher-order temporal network system

View Online

Export Citation

CrossMark

Junyuan Shi and Linhe Zhu[a)]

## AFFILIATIONS

School of Mathematical Sciences, Jiangsu University, Zhenjiang 212013, People's Republic of China

[a)]Author to whom correspondence should be addressed: zlhnuaa@126.com

## ABSTRACT

Reaction-diffusion processes on networked systems have received mounting attention in the past two decades, and the corresponding network dynamics theories have been continuously enriched with the advancement of network science. Recently, time-varying feature and many-body interactions have been discovered on various and numerous networks in real world like biological and social systems, and the study of contemporary network science has gradually moved away from the historically static network frameworks that are based on pairwise interactions. We are aimed to propose a general and rudimentary framework for Turing instability of reaction-diffusion processes on higher-order temporal networks. Firstly, we define a brand Laplacian to depict higher-order temporal diffusion behaviors on networks. Furthermore, the general form of higher-order temporal reaction-diffusion systems with frequency of oscillation is defined, and a time-independent and concise form is obtained by equivalent substitution and method of averaging. Next, we discuss the two cases of homogeneous and heterogeneous network systems and give the equivalent conditions of Turing instability through linear stability analysis. Finally, in the numerical simulation part, we verify and discuss the validity of the above theoretical framework and study the effect of the frequency of oscillation of higher-order temporal network on reaction-diffusion processes. Our study has revealed that higher-order temporal reaction-diffusion, which takes into account both time-varying feature and many-body interactions, can formulate innovative and diversified patterns. Moreover, these are significantly differences from patterns in continuous space and patterns on traditional networks.

## I. INTRODUCTION

Alan Turing, known as the pioneer of modern computing, introduced the concepts of the Turing machine and Turing test, which greatly influenced the development of computer science and artificial intelligence.[1] Interestingly, Turing played a key role in advancing the field of mathematical biology through his substantial contributions. More precisely, he pioneered a self-organizing mechanism to explain pattern-making in 1952 based on symmetry breaking instability in natural biological systems, which did not fully adhere to established physical principles such as mirror image symmetry.[2,3] Turing patterns are formed by introducing discontinuous diffusion into linear systems at a specific moment to destabilize homogeneous and stable systems. These nonlinear systems are referred to as reaction-diffusion systems and are widely used in research fields such as the pattern formulation.[4–11] Furthermore, as the understanding grows that many dynamic processes take place in discrete media rather than continuous space, the significance of network dynamics has become more pronounced and has undergone rapid progress in recent years.[12–17] Network-organized reaction-diffusion process in which species within patches or nodes interact with each other and species mobility in the diffusion process is achieved through structured links.[18–20] A general theoretical method for analyzing the stability of reaction-diffusion systems on network systems is proposed in 1971,[21] and since then the pattern formulation on different networks has become a hot research topic.[22–27] Notably, the features of complex networks have a significant influence on the dynamic behavior of the networked reaction-diffusion process, so we need to enrich and improve the theory of pattern formulation on various complex networks.

The study of complex networks with Erdős–Rényi (ER) networks[28–30] as origin, until Watts–Strogatz (WS) networks[31] and Barabási–Albert (BA) networks[32] pioneered to reveal the small-world nature and power-law degree distributions of networks have received high interesting and made unexpected development in the last two decades. For example, the use of network models to analyze intricate interactions between entities is the cornerstone of contemporary research on biological systems and has a wide range of applications such as neuroscience and human disease.[33] Moreover, some novel frameworks combining statistical physics and network science provide new ideas and insights for the study of complex Earth systems, providing reliable theories and predictions for scientific research and policy making in various fields such as climate change and extreme events.[34] In addition, the study of the emergence of urban scaling through mathematical models that take into account network structure and properties has also received widespread attention recently.[35] These recent authoritative reviews systematically organize the development of complex networks and their important results in other fields, which reflect two major trends in the development of complex networks. On the one hand, the dynamical behavior on modern complex networks, including chaos and bifurcation, has gradually become a critical issue in the future researches of network science. On the other hand, the higher-order feature and time-varying feature of complex networks have become more meaningful and will become the core of the future development of network science.

Recently, many-body interactions have been found in mounting scenarios in the real world. So one needs to use a higher-order view of complex networks and the dynamical behavior on them, rather than limiting oneself to using links to represent pairwise interactions of historical systems.[36–39] Dynamical processes on higher-order networks have likewise received a great deal of attention.[40] Contrary to traditional static network systems, temporal networks have a topology that changes over time.[41,42] Although not as simple to analyze and study as static networks, the advantages brought by the dynamic nature of temporal networks are still considered to be a more effective network in many areas of application.[43–45] Temporal network systems on dynamics has gradually become a hot research of great importance in recent times.[14,46,47]

Higher-order temporal networks synthesize the higher-order interaction and time-varying features of the real world, and have shown great potential for applications despite their recent emergence.[48] The dynamical behavior on higher-order temporal network systems has also been more widely and deeply studied.[49–52] For example, hypergraphs or simplicial complexes provide a way to describe the opinion dynamics among multiple individuals in a social system, rather than being limited to pairwise interactions. In addition to advances in topological structures, higher-order temporal networks better describe these time-varying interaction patterns from a temporal dimension, which are also common in social systems. In particular, communication methods such as the Internet have accelerated the transformation of network structures in social systems. Therefore, the reaction-diffusion process on higher-order temporal networks has the potential to be a novel approach to modeling opinion dynamics in social systems. However, the absence of Turing bifurcation theory for higher-order temporal networks hinders us from conducting a more comprehensive and systematic examination of networked systems and their dynamic processes. To address this deficiency, we propose a comprehensive and universal framework as the theoretical foundation for the corresponding investigations presented in this study.

In this paper, we make a contribution by introducing the concept of anchoring pattern formulation in higher-order temporal network systems for the first time. We begin by defining universal Laplacian operations and reaction-diffusion systems tailored specifically for higher-order temporal networks. Next, we propose a theoretical framework for linear stability analysis on both homogeneous and heterogeneous networks, respectively. Furthermore, we provide the equivalent conditions for the occurrence of Turing instability on these network systems. Additionally, we conduct extensive numerical simulations to validate the theoretical framework and facilitate a more comprehensive discussion. In addition, we pay special attention to heterogeneous networks characterized by different oscillation frequencies in order to demonstrate the soundness and effectiveness of the averaging method utilized in our analysis.

## II. DEFINITION OF HIGHER-ORDER TEMPORAL REACTION-DIFFUSION SYSTEMS

Firstly, we define the universal Laplacian operator of the higher-order temporal networked systems with frequency of oscillation $\mathfrak{f}$ by the incidence matrices.[50]

$$G(\mathfrak{f}T) = \{G(\mathfrak{f}t) = (V, S(\mathfrak{f}t)), t \in [0, T]\}$$

is a higher-order temporal network consisting of various simplices at different moments. In Fig. 1, a simplex complex consisting of 0-simplex, 1-simplex and 2-simplex is shown, with different interacting species on the network nodes.

$V = \{v_1, \ldots, v_N\}$ labels the $N$ nodes on the network and $N_{[m]}(\mathfrak{f}t)$ denotes the number of m-dimension simplices in $G(\mathfrak{f}t)$ [note that $N_{[0]} = N$ for all simplicial complexes $G(\mathfrak{f}t)$]. $S(\mathfrak{f}t) = \{\sum_{d=1}^{M} S_d(\mathfrak{f}t)\}$ where $S_d(\mathfrak{f}t) = \{d - \text{simplex} | d - \text{simplex} \in G(\mathfrak{f}t)\}$ is the set of $d$-simplices of $G(\mathfrak{f}t)$, $d = 1, \ldots, M$ with $M = \max\{d | d - \text{simplex} \in G(\mathfrak{f}T)\}$. A $d$-simplex $[v_0, v_1, \ldots, v_d]$ can be oriented by means of $\tau(\pi)$ that can react to the parity of the pernutation $\pi$:

$$[v_0, v_1, \ldots, v_d] = (-1)^{\tau(\pi)} [v_{\pi(0)}, v_{\pi(1)}, \ldots, v_{\pi(d)}]. \tag{1}$$

The $d$-chain $\mathscr{C}_d$ is composed of elements that belong to a free abelian group and the set of all oriented d-simplices $S_d(\mathfrak{f}t)$ in the simplex complex $G(\mathfrak{f}t)$ is the basis of its linear space. Therefore, we use the following linear combination of all oriented $d$-simplices denote its elements:

$$\gamma = [\mu_0, \mu_1, \ldots, \mu_m].$$

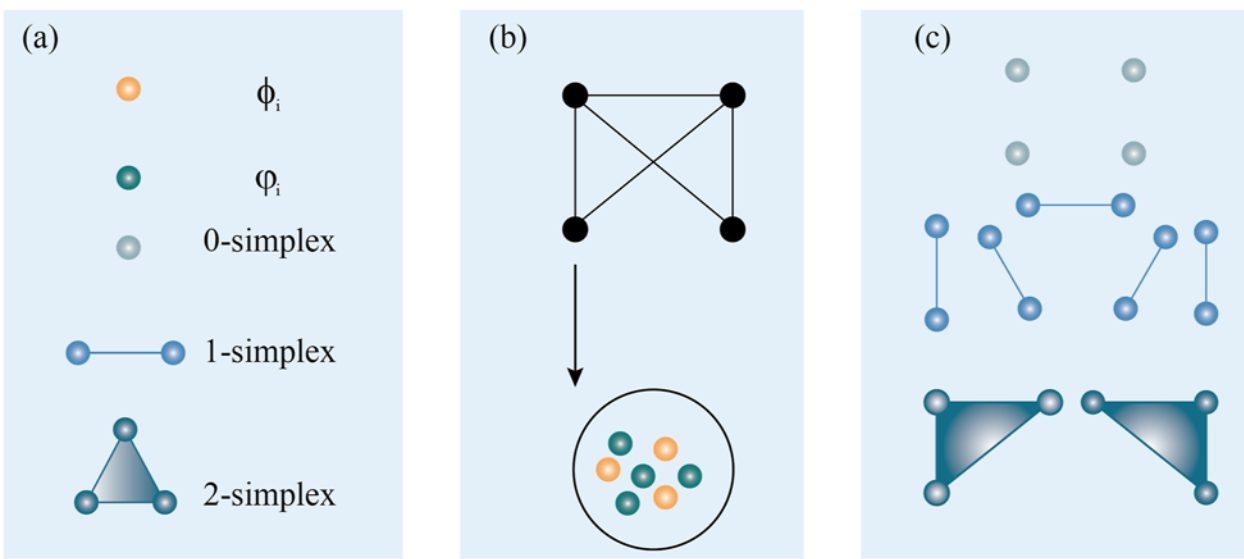

**FIG. 1.** Reaction-diffusion processes on higher-order temporal networks. (a) The representation of two species and d-simplices on higher-order temporal networks with metapopulations. This figure shows some common and typical d-simplex of simplicial complexes. (b) Two interactive species live on different nodes on networks. Each node generally includes two different species, which interact with each other within the node and have mobility on the network, which is closely related to the network structure. (c) The basic structure of a higher-order temporal network model in (b) is composed of simplices, which serve as representations of d-dimensional group interactions. For clarity and brevity, we only show some relatively low-order but common structures.

The boundary map $\partial_d : \mathscr{C}_d \to \mathscr{C}_{d-1}$ is a linear operator which maps an $d$-simplex to an $(d-1)$-chain consisting of the simplices and its boundary by the following way:

$$\partial_d[\mu_0, \mu_1, \ldots, \mu_m] = \sum_{r=0}^{m} (-1)^r [\mu_0, \mu_1, \ldots, \mu_{r-1}, \mu_{r+1}, \ldots, \mu_m]. \tag{2}$$

To this point, we introduce the oriented simplices, $d$-chain and boundary map as our mathematical tools for defining the incidence matrices $\mathbf{B}(\mathfrak{f}t)$, where $\mathfrak{f}$ labels the frequency of networks oscillations as mentioned previously. We use the ordered list $S_d(\mathfrak{f}t)$ as a set of basis for the linear space of m-chains, so that the $d$-boundary operator $\partial_d$ is related to the nonzero elements of the incidence matrix $\mathbf{B}_{[\mathbf{d}]}(\mathfrak{f}t)$ as follows:

$$\left[\mathbf{B}_{[\mathbf{d}]}(\mathfrak{f}t)\right]_{\gamma_2,\gamma_1} = (-1)^r, \tag{3}$$

where

$$\gamma_1 = [\mu_0, \mu_1, \ldots, \mu_m],$$
$$\gamma_2 = [\mu_0, \mu_1, \ldots, \mu_{r-1}, \mu_{r+1}, \ldots, \mu_m].$$

Finally, we define the higher-order Laplacian matrix $\mathscr{L}_{[\mathbf{d}]}(\mathfrak{f}t)$ according to the incidence matrix $\mathbf{B}_{[\mathbf{d}]}(\mathfrak{f}t)$, which describes the behaviors of diffusion between $d$-simplices. Noting that diffusion between $d$-simplices is always realized by a $(d+1)$-simplices or a $(d-1)$-simplices. Thus, the higher-order temporal Laplacian operator $\mathscr{L}_{[\mathbf{d}]}(\mathfrak{f}t)$, with its elements denoted as $l_{ij}(\mathfrak{f}t)$, where $i, \quad j = 0, 1, \ldots, N_{[d]}(\mathfrak{f}t)$, can perform a Hodge decomposition in the following form:

$$\mathscr{L}_{[\mathbf{d}]}(\mathfrak{f}t) = \mathscr{L}^{\mathbf{up}}_{[\mathbf{d}]}(\mathfrak{f}t) + \mathscr{L}^{\mathbf{down}}_{[\mathbf{d}]}(\mathfrak{f}t), \tag{4}$$

with

$$\mathscr{L}^{\mathbf{up}}_{[\mathbf{d}]}(\mathfrak{f}t) = \mathbf{B}_{[\mathbf{d+1}]}(\mathfrak{f}t)\ \mathbf{B}^{\mathbf{T}}_{[\mathbf{d+1}]}(\mathfrak{f}t),$$
$$\mathscr{L}^{\mathbf{down}}_{[\mathbf{d}]}(\mathfrak{f}t) = \mathbf{B}^{\mathbf{T}}_{[\mathbf{d}]}(\mathfrak{f}t)\ \mathbf{B}_{[\mathbf{d}]}(\mathfrak{f}t).$$

After giving the method of defining the higher-order Laplacian matrix, we use it to define the following higher-order reaction-diffusion system, which describes the reaction process between two interacting species $\phi_i$ and $\varphi_i$ within a node $i$ and their diffusion behaviors among d-simplices in the networks, $d = 1, \ldots, M$, and then equally divided among the various nodes involved:

$$\begin{aligned}
\frac{d\phi_i}{dt} &= f(\phi_i, \varphi_i) + \sum_{d=0}^{M} D_{\phi}^{[d]} \sum_{v_i \in S_k^{[d]}(\mathfrak{f}t)} \sum_{j=1}^{N_{[d]}(\mathfrak{f}t)} l_{kj}^{[d]}(\mathfrak{f}t) \langle \phi_{S_j^{[d]}} \rangle, \\
\frac{d\varphi_i}{dt} &= g(\phi_i, \varphi_i) + \sum_{d=0}^{M} D_{\varphi}^{[d]} \sum_{v_i \in S_k^{[d]}(\mathfrak{f}t)} \sum_{j=1}^{N_{[d]}(\mathfrak{f}t)} l_{kj}^{[d]}(\mathfrak{f}t) \langle \varphi_{S_j^{[d]}} \rangle,
\end{aligned} \tag{5}$$

where $f$ and $g$ are the generic nonlinear reaction functions, $D_{\phi}^{[d]}$ and $D_{\varphi}^{[d]}$ label the diffusion coefficients of the two species on d-simplies respectively, $S_k^{[d]}(\mathfrak{f}t)$ donates the $k$-th simplex in the set $S_d(\mathfrak{f}t)$, $\langle \phi_{S_j^{[d]}} \rangle$ and $\langle \varphi_{S_j^{[d]}} \rangle$ denote the quantities of two simplices that are evenly distributed to node $i$ after being diffused from other d-simplices to $S_k^{[d]}(t)$.

It is not easy to perform a linear stability analysis of the above reaction-diffusion system (5) with complex diffusion terms that leading to different dimension of the temporal Laplacian, so we get a equivalent matrix $\mathbf{C}$ to come by the following two steps. $C_{ij}^{[d]}(\mathfrak{f}t)$ is equal to 0 if $i$ is not adjacent to $j$ and there exactly exist some constants that satisfy the following equation if the degrees of each node is more than 1, which is always true for a complex network:

$$\begin{aligned} \sum_{v_i \in S_k^{[d]}(\mathfrak{f}t)} \sum_{j=1}^{N_{[d]}(\mathfrak{f}t)} l_{kj}^{[d]}(\mathfrak{f}t)\langle \phi_{S_j^{[d]}} \rangle &= \sum_{j=1}^{N} C_{ij}^{[d]}(\mathfrak{f}t)\phi_j, \\ \sum_{v_i \in S_k^{[d]}(\mathfrak{f}t)} \sum_{j=1}^{N_{[d]}(\mathfrak{f}t)} l_{kj}^{[d]}(\mathfrak{f}t)\langle \varphi_{S_j^{[d]}} \rangle &= \sum_{j=1}^{N} C_{ij}^{[d]}(\mathfrak{f}t)\varphi_j. \end{aligned} \tag{6}$$

The equivalent matrix $\mathbf{C}^{[\mathbf{d}]}(\mathfrak{f}t)$ is a time dependent and asymmetric matrix whose elements are $C_{ij}^{[d]}(\mathfrak{f}t)$. The Eq. (6) intuitively seems infeasible, but the time-varying property guarantees the existence of $\mathbf{C}(\mathfrak{f}\mathbf{t})$ throughout $(0, T)$. Therefore, we can simplify the previous reaction-diffusion system (5) according to (6):

$$\begin{aligned} \frac{d\phi_i}{dt} &= f(\phi_i, \varphi_i) + \sum_{d=0}^{M} D_\phi^{[d]} \sum_{j=1}^{N} C_{ij}^{[d]}(\mathfrak{f}t)\phi_j, \\ \frac{d\varphi_i}{dt} &= g(\phi_i, \varphi_i) + \sum_{d=0}^{M} D_\varphi^{[d]} \sum_{j=1}^{N} C_{ij}^{[d]}(\mathfrak{f}t)\varphi_j. \end{aligned} \tag{7}$$

## III. THE THEORY OF PATTERN FORMULATION ON HIGHER-ORDER TEMPORAL NETWORK SYSTEMS

The system (7) can describe the reaction-diffusion processes on higher-order time-varying networks accurately and we are first aimed to find a simple and time-independent system. In the following, in order to study the dynamical behavior of system (5) on networks more deeply and comprehensively, we propose the theoretical framework of pattern formulation on higher-order temporal networked systems.

Firstly, image that system (7) exists an homogeneous and stable fixed point $(\phi^*, \varphi^*)$, such that $f(\phi^*, \varphi^*) = g(\phi^*, \varphi^*) = 0$, satisfies both $\det(\mathbf{J}) > 0$ and $\mathrm{tr}(\mathbf{J}) < 0$ with $\mathbf{J} = \frac{\partial(f, g)}{\partial(\phi^*, \varphi^*)}$ is the Jacobian matrix of the nonlinear reaction functions, whose elements are $J_{ij}, i, j = 1, 2$. We introduce an inhomogeneous perturbation $\vec{\imath} = (\delta\phi_i, \delta\varphi_i)^T = (\phi_i - \phi^*, \varphi_i - \varphi^*)^T$ that destabilizes the homogeneous fixed point $(\phi^*, \varphi^*)$ to cause Turing pattern arise. Defining the perturbation vector $\vec{\zeta}(\mathfrak{f}t) = (\delta\phi_1, \delta\varphi_1, \ldots, \delta\phi_N, \delta\varphi_N)^T$ and linearizing this vector around $(\phi^*, \varphi^*)$ to obtain the following equation:

$$\frac{d\vec{\zeta}(\mathfrak{f}t)}{dt} = \frac{1}{\mathfrak{f}}\left(\mathbf{I}_N \otimes \mathbf{J} + \sum_{d=0}^{M} \mathbf{C}^{[\mathbf{d}]}(\mathfrak{f}t) \otimes \mathbf{D}^{[\mathbf{d}]}\right)\vec{\zeta}(\mathfrak{f}t), \tag{8}$$

where $\mathbf{I}_N$ is $N \times N$ identity matrix and $\mathbf{D}^{[\mathbf{d}]} = \begin{bmatrix} D_\phi^{[d]} & 0 \\ 0 & D_\varphi^{[d]} \end{bmatrix}$ labels the diffusion matrix respecting to d-simplices on the networks. We define the following time-independent matrix $\mathbf{C}^{[\mathbf{d}]}$:

$$\mathbf{C}^{[\mathbf{d}]} = \lim_{T \to \infty} \frac{1}{T} \int_0^T \mathbf{C}^{[\mathbf{d}]}(\mathfrak{f}t)dt. \tag{9}$$

Replacing the equivalent matrix $\mathbf{C}^{[\mathbf{d}]}(\mathfrak{f}t)$ in the time-dependent reaction-diffusion system (7) with the averaging equivalent matrix $\mathbf{C}^{[\mathbf{d}]}$, to obtain the following time-independent reaction-diffusion system:

$$\begin{aligned} \frac{d\phi_i}{dt} &= f(\phi_i, \varphi_i) + \sum_{d=0}^{M} D_\phi^{[d]} \sum_{j=1}^{N} C_{ij}^{[d]}\phi_j, \\ \frac{d\varphi_i}{dt} &= g(\phi_i, \varphi_i) + \sum_{d=0}^{M} D_\varphi^{[d]} \sum_{j=1}^{N} C_{ij}^{[d]}\varphi_j. \end{aligned} \tag{10}$$

In the following, we first discuss the closeness between time-independent reaction-diffusion system (10) and time-dependent reaction-diffusion system (7) with high frequency of network oscillation, and then gives a theory of Turing instability based on the former. The validity of this time-independent theoretical framework for higher-order temporal networked systems is guaranteed by the method of averaging.

Method of averaging is an important way to approximate dynamical systems with time-scales as time-independent reaction-diffusion systems. Lagrange used the method of averaging[53,54] to transform the gravitational three-body problem into a perturbation of the two-body problem in 1788 and the validity of this method was verified by a number of academic rigor in 1928. Subsequently, this method has been used as a fundamental approach to the study of nonlinear oscillations and has solved a number of important problems, including the Krylov–Bogoliubov. Applying method of averaging, we have

$$\begin{aligned}\frac{d\vec{\xi}}{dt} &= \frac{1}{\mathfrak{f}T}\int_0^T \left(\mathbf{I}_N \otimes \mathbf{J} + \sum_{d=0}^{M} \mathbf{C}^{[\mathbf{d}]}(\mathfrak{f}t) \otimes \mathbf{D}^{[\mathbf{d}]}\right) dt\vec{\xi} \\ &= \frac{1}{\mathfrak{f}}\left(\mathbf{I}_N \otimes \mathbf{J} + \sum_{d=0}^{M} \mathbf{C}^{[\mathbf{d}]} \otimes \mathbf{D}^{[\mathbf{d}]}\right)\vec{\xi},\end{aligned} \tag{11}$$

with

$$\vec{\xi}(\mathfrak{f}t) - \vec{\zeta}(\mathfrak{f}t) = O(1/\mathfrak{f}), \tag{12}$$

where $T$ is the period of the transformation period of the network, and the above conclusions can be generalized to non-periodic network systems when $T$ tends to infinity.

The emergence of obvious heterogeneity in a variety of network systems has driven the researches of dynamical behavior on heterogeneous networks.[55–59] Unlike the homogeneous case, where different species within the same node still interact with each other, on heterogeneous networks the diffusion processes of different species occur in distinct diffusion environments, characterized by diverse network structures (see Fig. 2). This leads to a more diverse range of pattern formulation mechanisms. We also present a comprehensive heterogeneous higher-order temporal theoretical framework for pattern formation in network systems. Firstly, we similarly assume that $(\phi^*, \varphi^*)$ is a stable and homogeneous fixed point and introduce a perturbation vector $\vec{\chi} = (\phi_1 - \phi^*, \varphi_1 - \varphi^*, \ldots, \phi_N - \phi^*, \varphi_N - \varphi^*)$ to the following reaction-diffusion system on the heterogeneous network, which consists of a process similar to that of the homogeneous case process obtained.

$$\begin{aligned}\frac{d\phi_i}{dt} &= f(\phi_i, \varphi_i) + \sum_{d=0}^{M} D_\phi^{[d]} \sum_{j=1}^{N} M_{ij}^{[d]} \phi_j, \\ \frac{d\varphi_i}{dt} &= g(\phi_i, \varphi_i) + \sum_{d=0}^{M} D_\varphi^{[d]} \sum_{j=1}^{N} M_{ij}^{[d]} \varphi_j,\end{aligned} \tag{13}$$

subject to

$$\begin{aligned}&\mathbf{M}^{[\mathbf{d}]} = \lim_{T\to\infty} \frac{1}{T}\int_0^T \mathbf{M}^{[\mathbf{d}]}(\mathfrak{f}t)dt, \\ &\sum_{v_i \in S_k^{[d]}(\mathfrak{f}_1 t)} \sum_{j=1}^{N_{[d]}(\mathfrak{f}_1 t)} A_{kj}^{[d]}(\mathfrak{f}_1 t)\langle \phi_{S_j^{[d]}} \rangle = \sum_{j=1}^{N} M_{ij}^{[d]}(\mathfrak{f}t)\phi_j, \\ &\sum_{v_i \in S_k^{[d]}(\mathfrak{f}_2 t)} \sum_{j=1}^{N_{[d]}(\mathfrak{f}_2 t)} B_{kj}^{[d]}(\mathfrak{f}_2 t)\langle \varphi_{S_j^{[d]}} \rangle = \sum_{j=1}^{N} M_{ij}^{[d]}(\mathfrak{f}t)\varphi_j.\end{aligned}$$

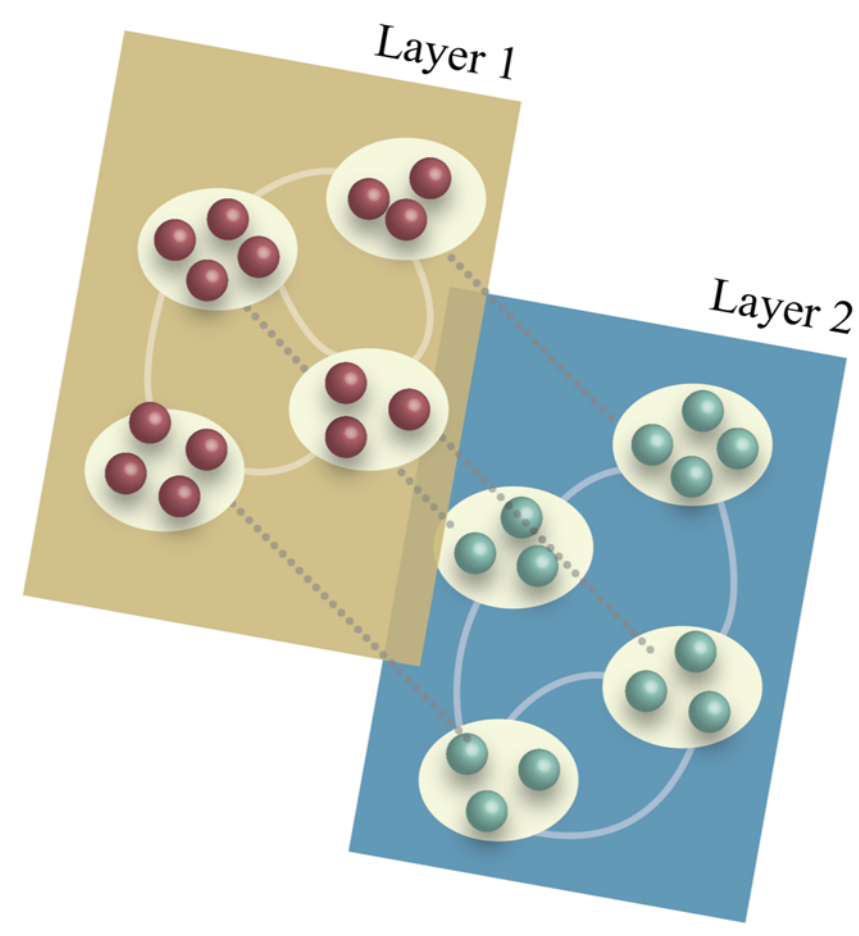

**FIG. 2.** Schematic representation of a heterogeneous network system with two different species. Layer 1 and layer 2 label the diffusion environments of the two species.

$\mathfrak{f} = \max(\mathfrak{f}_1, \mathfrak{f}_2)$, $\mathfrak{f}_1, \mathfrak{f}_2$ are the oscillation frequencies of the Laplace operators $\mathbf{A}$ and $\mathbf{B}$ corresponding to the two different networked systems. Furthermore, by linearizing perturbation vector $\vec{\chi}$ around the fixed point $(\phi^*, \varphi^*)$, we get the following equation:

$$\frac{d\vec{\chi}}{dt} = \frac{1}{\mathfrak{f}}\left(\mathbf{I}_N \otimes \mathbf{J} + \sum_{d=0}^{M} \mathbf{M}^{[\mathbf{d}]} \otimes \mathbf{D}^{[\mathbf{d}]}\right)\vec{\chi}. \tag{14}$$

It is then possible to go through the same process and we arrive at a theory of pattern formulation that applies to both the linearized system (8) and the (14). The small perturbations $(\delta\phi_i, \delta\varphi_i)$ are expanded as:

$$\begin{pmatrix} \delta\phi_i \\ \delta\varphi_i \end{pmatrix} = \begin{pmatrix} C_{\phi}^{[d,\alpha]} \\ C_{\varphi}^{[d,\alpha]} \end{pmatrix} e^{\lambda_\alpha t} v_i^{[d,\alpha]}, \quad \alpha = 1, \ldots, N, \tag{15}$$

where the constants $C_{\phi}^{[d,\alpha]}$ and $C_{\varphi}^{[d,\alpha]}$ are determined by the initial conditions of the systems and $v_i^{[d,\alpha]}$ is the $i$-th element of the eigenvector $v^{[d,\alpha]}$ of the $\mathbf{C}^{[\mathbf{d}]}$. The complex number $\Lambda_\alpha^{[d]} = r_\alpha^{[d]} \exp(i\theta_\alpha^{[d]}) = r_\alpha^{[d]}(\cos\theta_\alpha^{[d]} + i\sin\theta_\alpha^{[d]})$ is specified as the eigenvalue corresponding to $v^{[d,\alpha]}$ of $\mathbf{C}^{[\mathbf{d}]}$ with the following relation:

$$\sum_{j=1}^{N} C_{ij}^{[d]} v_j^{[d,\alpha]} = \Lambda_\alpha^{[d]} v_i^{[d,\alpha]}. \tag{16}$$

In the follow-up of our work, $(\cdot)_{\mathrm{Re}}$ and $(\cdot)_{\mathrm{Im}}$ label the real and imaginary parts of the bracketed portions, respectively. We can obtain $N$ linear approximation systems of size $2 \times 2$ by projecting (8) and inserting the above two equations into it, whose characteristic equation is

$$\det(\lambda_\alpha \mathbf{I_2} - \mathbf{J}^{[\boldsymbol{\alpha}]}) = 0 \quad \text{with} \quad \mathbf{J}^{[\boldsymbol{\alpha}]} = \mathbf{J} + \sum_{d=0}^{M} \Lambda_\alpha^{[d]} \mathbf{D}^{[\mathbf{d}]}, \tag{17}$$

subject to

$$\begin{aligned}
&\left(\mathrm{tr}\mathbf{J}^{[\boldsymbol{\alpha}]}\right)_{\mathrm{Re}} = \mathrm{tr}\mathbf{J} + \sum_{d=1}^{M} r_\alpha^{[d]} \cos\theta_\alpha^{[d]} \mathrm{tr}\mathbf{D}^{[\mathbf{d}]}, \\
&\left(\mathrm{tr}\mathbf{J}^{[\boldsymbol{\alpha}]}\right)_{\mathrm{Im}} = \sum_{d=1}^{M} r_\alpha^{[d]} \sin\theta_\alpha^{[d]} \mathrm{tr}\mathbf{D}^{[\mathbf{d}]}, \\
&\left(\det\mathbf{J}^{[\boldsymbol{\alpha}]}\right)_{\mathrm{Re}} = \det\mathbf{J} + J_{11}\sum_{d=1}^{M} D_{\varphi}^{[d]} r_\alpha^{[d]} \cos\theta_\alpha^{[d]} \\
&+ J_{22}\sum_{d=1}^{M} D_{\phi}^{[d]} r_\alpha^{[d]} \cos\theta_\alpha^{[d]} + \sum_{d=1}^{M} D_{\phi}^{[d]} r_\alpha^{[d]} \cos\theta_\alpha^{[d]} \sum_{d=1}^{M} D_{\varphi}^{[d]} r_\alpha^{[d]} \cos\theta_\alpha^{[d]} \\
&- \sum_{d=1}^{M} D_{\phi}^{[d]} r_\alpha^{[d]} \sin\theta_\alpha^{[d]} \sum_{d=1}^{M} D_{\varphi}^{[d]} r_\alpha^{[d]} \sin\theta_\alpha^{[d]}, \\
&\left(\det\mathbf{J}^{[\boldsymbol{\alpha}]}\right)_{\mathrm{Im}} = J_{11}\sum_{d=1}^{M} D_{\varphi}^{[d]} r_\alpha^{[d]} \sin\theta_\alpha^{[d]} + J_{22}\sum_{d=1}^{M} D_{\phi}^{[d]} r_\alpha^{[d]} \sin\theta_\alpha^{[d]} \\
&+ \sum_{d=1}^{M} D_{\phi}^{[d]} r_\alpha^{[d]} \sin\theta_\alpha^{[d]} \sum_{d=1}^{M} D_{\varphi}^{[d]} r_\alpha^{[d]} \cos\theta_\alpha^{[d]} \\
&+ \sum_{d=1}^{M} D_{\varphi}^{[d]} r_\alpha^{[d]} \sin\theta_\alpha^{[d]} \sum_{d=1}^{M} D_{\phi}^{[d]} r_\alpha^{[d]} \cos\theta_\alpha^{[d]}.
\end{aligned}$$

In addition, for the sake of brevity of presentation, we denote the real and imaginary part of $\sqrt{(\mathrm{tr}\mathbf{J}^{[\boldsymbol{\alpha}]})^2 - 4\det\mathbf{J}^{[\boldsymbol{\alpha}]}}$ as R and I. Moreover, by the characteristic Eq. (17) and the formula for the square root of a complex number to solve for $\lambda_\alpha$ of the form:

$$\begin{aligned}
2\lambda_\alpha &= \mathrm{tr}\mathbf{J}^{[\boldsymbol{\alpha}]} + \sqrt{(\mathrm{tr}\mathbf{J}^{[\boldsymbol{\alpha}]})^2 - 4\det\mathbf{J}^{[\boldsymbol{\alpha}]}} \\
&= \left[\left(\mathrm{tr}\mathbf{J}^{[\boldsymbol{\alpha}]}\right)_{\mathrm{Re}} + \mathrm{R}\right] + i\left[\left(\mathrm{tr}\mathbf{J}^{[\boldsymbol{\alpha}]}\right)_{\mathrm{Im}} + \mathrm{I}\right],
\end{aligned} \tag{18}$$

where

$$\mathrm{R} = \sqrt{\frac{\mathscr{X} + \mathscr{Z}}{2}} \quad \text{and} \quad \mathrm{I} = \frac{\mathscr{Y}}{|\mathscr{Y}|}\sqrt{\frac{\mathscr{Z} - \mathscr{X}}{2}},$$

subject to

$$\begin{aligned}
\mathscr{X} &= \left(\mathrm{tr}\mathbf{J}^{[\boldsymbol{\alpha}]}\right)_{\mathrm{Re}}^{2} - \left(\mathrm{tr}\mathbf{J}^{[\boldsymbol{\alpha}]}\right)_{\mathrm{Im}}^{2} - 4\left(\det \mathbf{J}^{[\boldsymbol{\alpha}]}\right)_{\mathrm{Re}},\\
\mathscr{Y} &= 2\left(\mathrm{tr}\mathbf{J}^{[\boldsymbol{\alpha}]}\right)_{\mathrm{Re}}\left(\mathrm{tr}\mathbf{J}^{[\boldsymbol{\alpha}]}\right)_{\mathrm{Im}} - 4\left(\det \mathbf{J}^{[\boldsymbol{\alpha}]}\right)_{\mathrm{Im}},\\
\mathscr{Z} &= \sqrt{\mathscr{X}^2 + \mathscr{Y}^2}.
\end{aligned}$$

$(\lambda_\alpha)_{\mathrm{Re}} = 0$ for some $\alpha = \alpha^*$ is the threshold for instability and the $\alpha$-th node become unstable when the linear growth rate $\lambda_\alpha$ has a positive real part. Furthermore, the Turing pattern arise as these unstable critical nodes grow while others with $(\lambda_\alpha)_{\mathrm{Re}} < 0$ keep stable. To demonstrate the Turing instability, it is equivalent to meet one of the following two scenarios, according to the Eq. (18):

$$\left(\mathrm{tr}\mathbf{J}^{[\boldsymbol{\alpha}]}\right)_{\mathrm{Re}} > 0 \quad \text{or} \tag{19}$$

$$\begin{aligned}
&\left(\mathrm{tr}\mathbf{J}^{[\boldsymbol{\alpha}]}\right)_{\mathrm{Re}} \leq 0 \quad \text{and} \quad \left(\det \mathbf{J}^{[\boldsymbol{\alpha}]}\right)_{\mathrm{Im}}^{2} < \left(\mathrm{tr}\mathbf{J}^{[\boldsymbol{\alpha}]}\right)_{\mathrm{Re}}\\
&\times \left[\left(\det \mathbf{J}^{[\boldsymbol{\alpha}]}\right)_{\mathrm{Re}}\left(\mathrm{tr}\mathbf{J}^{[\boldsymbol{\alpha}]}\right)_{\mathrm{Re}} + \left(\det \mathbf{J}^{[\boldsymbol{\alpha}]}\right)_{\mathrm{Im}}\left(\mathrm{tr}\mathbf{J}^{[\boldsymbol{\alpha}]}\right)_{\mathrm{Im}}\right].
\end{aligned} \tag{20}$$

## IV. NUMERICAL RESULTS

We shall here validate and discuss the above universal theoretical framework in the context of a specific case, which can be widely applied to various real-world scenarios such as rumor spreading. Contemporary rumor spreading is often realized in rapidly transforming social networks with distinctive community structures of complex networks,[60] which means that time-varying feature and many-body interactions become an increasingly non-negligible factor. A Suspicious-Infected (SI) reaction-diffusion model is developed to study the dynamics of rumor spreading on the complex networks in this work.[61] Furthermore, epidemic-spreading dynamics on temporal networks and higher-order networks is a very hot challenging to be studied.[41,62] We use the same reaction terms and combine them with both the time-varying feature and many-body interactions of contemporary social networks to build the following model for subsequent extensive numerical simulations.

$$\begin{aligned}
\frac{dS_i}{dt} &= rS_i\left(1 - \frac{S_i}{K}\right)\left(\frac{S_i}{A} - 1\right) - \beta S_i I_i + \sum_{d=0}^{M} D_S^{[d]} \sum_{j=1}^{N} C_{ij}^{[d]} S_j,\\
\frac{dI_i}{dt} &= \beta S_i I_i - \mu I_i^2 + \sum_{d=0}^{M} D_I^{[d]} \sum_{j=1}^{N} C_{ij}^{[d]} I_j.
\end{aligned} \tag{21}$$

We then demonstrate our theory by the reaction-diffusion model (21) on a higher-order temporal network system containing four nodes as shown in Fig. 3. Noting that the topology of the networks is encoded by higher-order temporal Laplacian. This network has four possible states denoted as Network 1–4 [see Fig. 3(a)] and may be in different network states as time evolves. In Fig. 3(b) the two-species heterogeneous case is shown, and the two blue dots of different color depths label the diffusion environments in which the two species are at different moments.

In the following numerical simulations, we first set the free parameters of the nonlinear reaction functions to values that stabilize the fixed point when no diffusion occurs, and then introduce diffusion between related nodes through positive diffusion coefficients to induce Turing instability. Furthermore, we verify the validity of our derived conditions for Turing instability on higher-order temporal network systems and discuss the influence of the frequency of network oscillation $\mathfrak{f}$ to our theory.

### A. Turing instability on homogeneous higher-order temporal network systems

In this case, two different species diffuse on the network whose frequency of oscillation is 10 as shown in Fig. 3. The model parameters are set as Figs. 4 and 5 corresponding to the two cases of conditions (19) and (20), respectively. The almond dots and lavender diamonds represent the traces and determinants of $\mathbf{J}^{[\boldsymbol{\alpha}]}$ at different moments. Moreover, we use the royal blue pentagram labels $\left(\mathrm{tr}\mathbf{J}^{[\boldsymbol{\alpha}]}\right)_{\mathrm{Re}}$ and orange pentagram labels $\left(\mathrm{tr}\mathbf{J}^{[\boldsymbol{\alpha}]}\right)_{\mathrm{Re}}\left[\left(\det \mathbf{J}^{[\boldsymbol{\alpha}]}\right)_{\mathrm{Re}}\left(\mathrm{tr}\mathbf{J}^{[\boldsymbol{\alpha}]}\right)_{\mathrm{Re}} + \left(\det \mathbf{J}^{[\boldsymbol{\alpha}]}\right)_{\mathrm{Im}}\left(\mathrm{tr}\mathbf{J}^{[\boldsymbol{\alpha}]}\right)_{\mathrm{Im}}\right] - \left(\det \mathbf{J}^{[\boldsymbol{\alpha}]}\right)_{\mathrm{Im}}^{2}$. Without affecting the result positively or negatively, we have multiplied the above two values by a positive constant for aesthetic purposes when visualizing. Next, Figs. 4(a) and 5(a) show cases where conditions (19) and (20) hold, respectively. Figures 4(b) and 5(b) show the phenomenon that the density of susceptible individuals in the four nodes of the higher-order temporal network changes over time for two different sets of parameters. A more pronounced fluctuation in the interval $t \in [0, 80]$ occurs in Fig. 5(b), but after this both converge to another steady state again. Figures 4(c) and 5(c) are Turing patterns that are arise on higher-order temporal network systems in distinct moments.

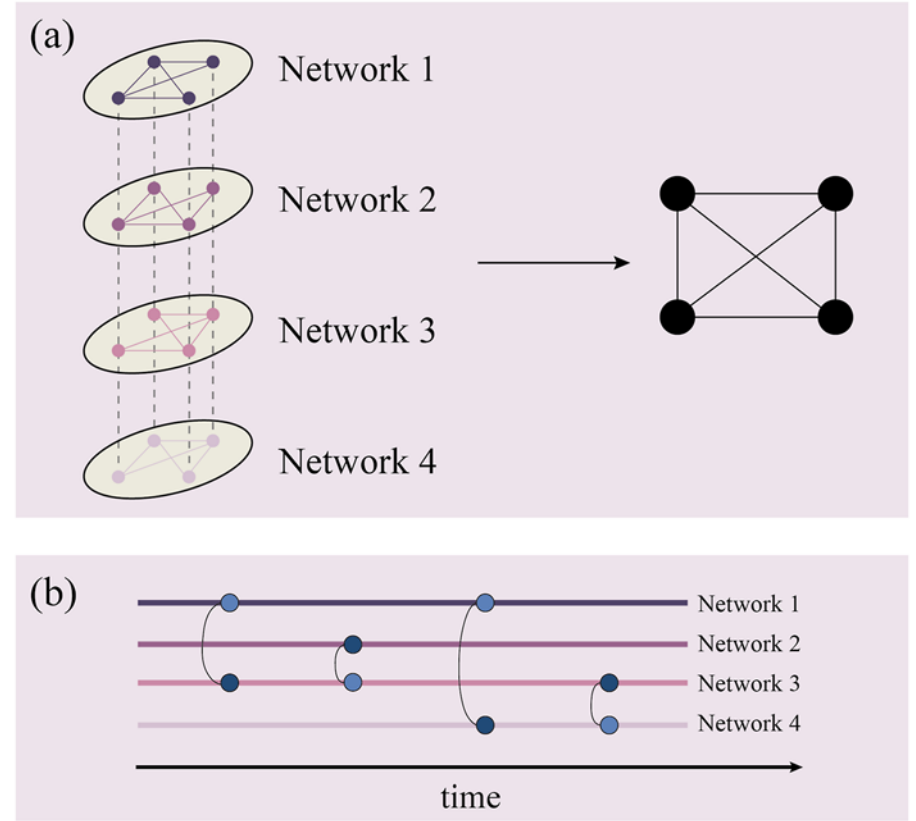

**FIG. 3.** A higher-order temporal network with four possible network configurations is depicted. (a) illustrates that these four possible networks can be averaged to form a static network that includes all the observed edges. These edges have different weights, which is related to the evolution of network weights. (b) At different time points, two species may be situated in different diffusion environments.

### B. Turing instability on heterogeneous higher-order temporal network systems

We now turn to the heterogeneous case, and in this section we demonstrate the validity and accuracy of the method of averaging in dealing with systems with time-scale by changing the frequency of network oscillation, which plays a very critical role in this functional method. We continue to discuss the case of the four nodes described above, but for clearer simulation effects, we generate random links between the four nodes to simulate some more universal networks. From the numerical results of Fig. 6 we find the following important phenomena. First, when the frequency $\mathfrak{f}_1$ of the diffusion environment of the susceptible individuals is 0.01, the slow oscillation of the network leads to significant fluctuations in the density of the susceptible individuals even in the case of the infected individuals with a frequency $\mathfrak{f}_2 = 100$ of the transformation of the diffusion environment. As the frequency $\mathfrak{f}_1$ rises to 1 and 100, the fluctuation rapidly decreases and flattens out to the situation in the average state. Moreover, this fluctuation flattens out further as $\mathfrak{f}_2$ increases. This illustrates the feasibility and accuracy of approximating time-varying systems with a time-independent averaging systems at a suitable frequency, for which there are no overly demanding requirements. Method of averaging has played an important role in a number of studies throughout history and has great potential to become a fundamental theoretical approach to subsequent studies of network dynamics with time-scale.

## V. DISCUSSION

Many-body interactions and time-varying feature have been found to play an mounting significant role in the recent extensive research, which determines the development of network science and the dynamical researches based on it in the direction of higher-order and time-varying. For example, Di Gaetano *et al.* provided analytical insights into the main topological properties of time-integrated hypergraphs and estimate percolation times in different classes of hypergraphs, highlighting the underestimation of percolation time when neglecting the higher-order nature of empirical social interactions.[52] Under this trend, it has become a critical issue to go beyond the traditional framework to establish and improve the theories of higher-order temporal networked systems and their corresponding network dynamics from a brand perspective.

In this work, we mainly focus on the Turing pattern on the network systems and propose a universal framework, including the way to establish reaction-diffusion systems and the general process of linear stability analysis to them. We first find that different parameters can cause Turing instability to satisfy the condition (19) or (20). These two arguments are closely related to the determinant and trace of the modified Jacobian $\mathbf{J}^{[\alpha]}$, which is consistent with the conclusion we have derived. Moreover, the extensive numerical results on the frequency of network oscillation $\mathfrak{f}$, which plays a critical role in the method of averaging, show that the increase of this parameter effectively averages out the apparent fluctuations, allowing higher-order temporal networked systems to be approximated as the averaging one. From this framework, we find some important properties of the higher-order temporal network systems and the reaction-diffusion processes in higher-order temporal network systems. Specially, the phenomena of Turing instability on higher-order temporal networked systems is observed for the first time. This presents a feasible idea and a universal theoretical framework for subsequent researches that are related to Turing theory or higher-order temporal. Remarkably, our theory can be viewed as a general case for the theory of pattern formulation on higher-order networks or temporal network. Time-varying many-body interactions profoundly affect dynamic behavior in complex systems, which has prompted people to introduce higher-order topological structures and time-varying networks into the study of fields such as climate systems and social systems. For example, we can establish a higher-order temporal network model and a corresponding reaction-diffusion system to describe climate change on a complex Earth system, and study the steady-state transition and tipping point of this dynamic system. This is particularly important for analyzing and predicting increasingly complex and severe climate phenomena. This theoretical framework is also useful

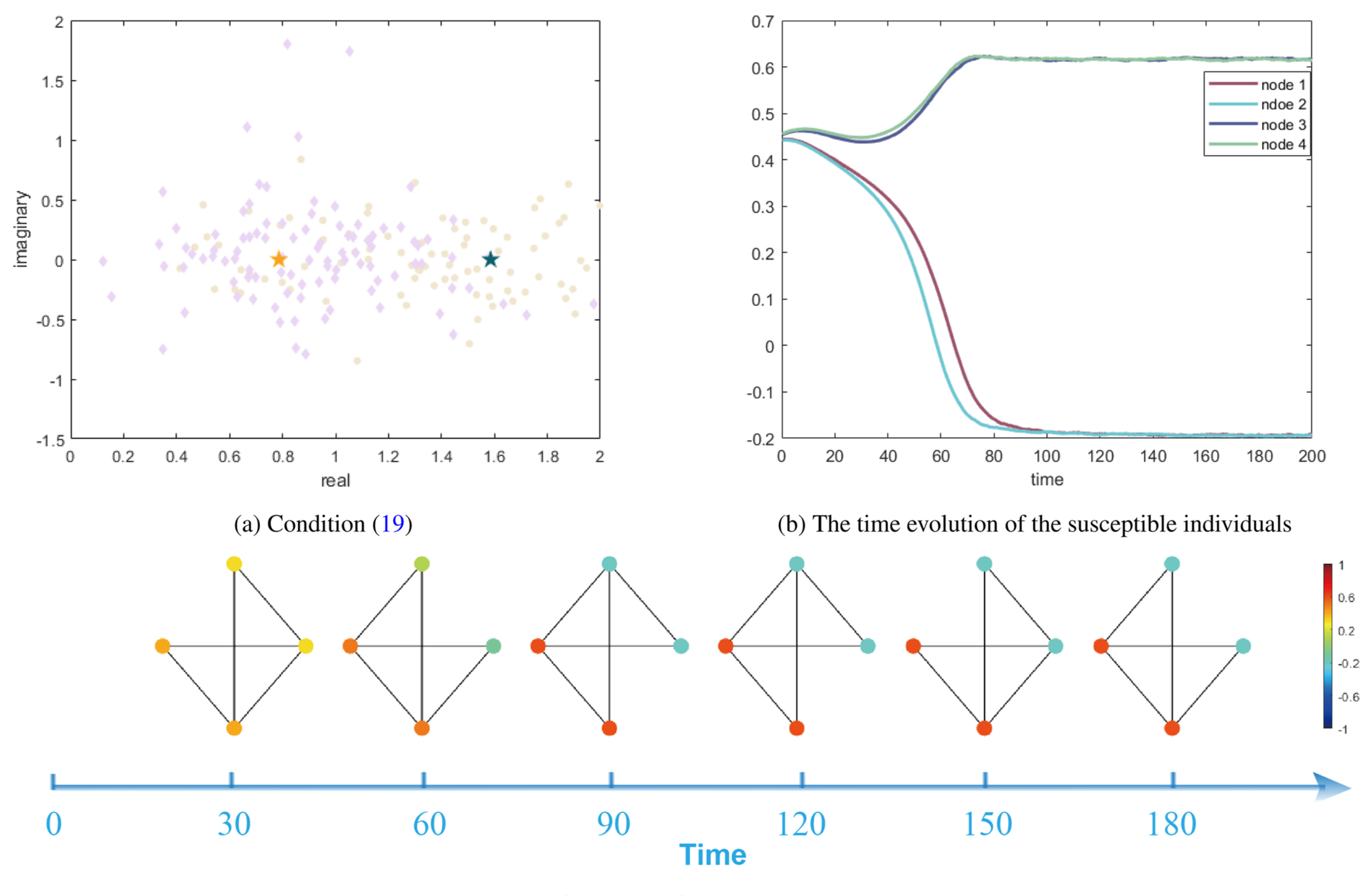

(a) Condition (19)

(b) The time evolution of the susceptible individuals

(c) The pattern formation of the susceptible individuals

**FIG. 4.** Turing instability on higher-order temporal network for condition (19). The reaction parameters are $K = 1$, $A = 0.08$, $\beta = 0.49$, $\mu = 0.53$, $r = 0.92$, the diffusion parameters are listed as follow: $D_S^{[1]} = 0.88$, $D_S^{[2]} = 0.001$, $D_S^{[3]} = 0.08$; $D_I^{[1]} = 0.11$, $D_I^{[2]} = 0.003$, $D_I^{[3]} = 0.07$. A statement elucidates that while many-body interactions hold significant importance, direct interactions between nodes continue to play a predominant role in certain scenarios.

for modeling social systems, which can also consider the impact of human behavior on climate and further develop coupled socio-climate models.[63]

## ACKNOWLEDGMENTS

This work is supported by China Postdoctoral Science Foundation (Grant No. 2023M731382).

## AUTHOR DECLARATIONS

### Conflict of Interest

The authors have no conflicts to disclose.

### Author Contributions

**Junyuan Shi**: Data curation (lead); Formal analysis (equal); Writing – original draft (lead). **Linhe Zhu**: Conceptualization (lead); Formal analysis (equal); Writing – review & editing (lead).

## DATA AVAILABILITY

The data that support the findings of this study are available in the section of the numerical results.

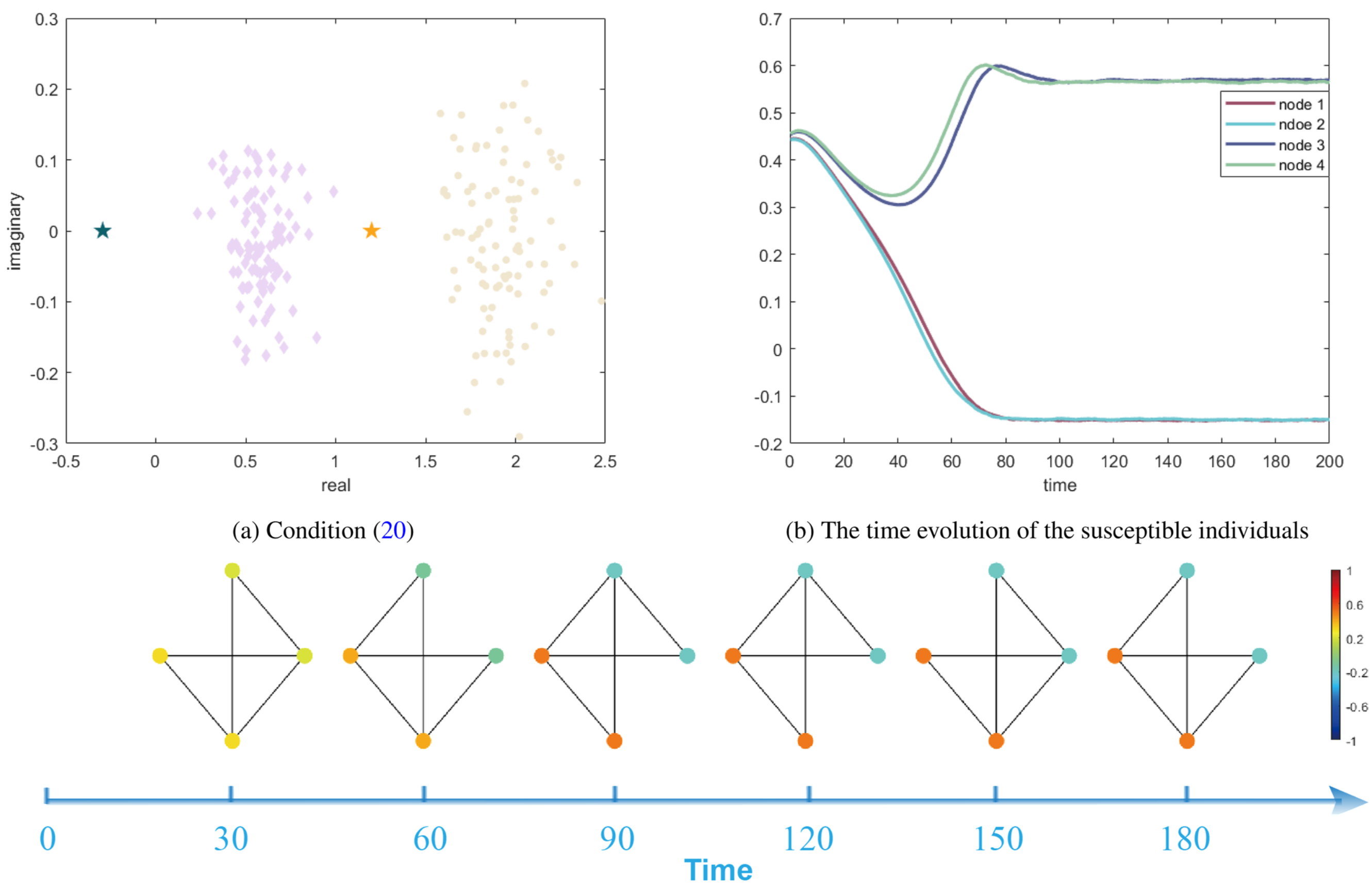

(a) Condition (20)

(b) The time evolution of the susceptible individuals

(c) The pattern formation of the susceptible individuals

**FIG. 5.** Turing instability on higher-order temporal network for condition (20). The reaction parameters remain consistent with Fig. 4. The diffusion parameters are listed as follow: $D_S^{[1]} = 0.81$, $D_S^{[2]} = 0.008$, $D_S^{[3]} = 0.08$; $D_I^{[1]} = 0.02$, $D_I^{[2]} = 0.082$, $D_I^{[3]} = 0.07$.

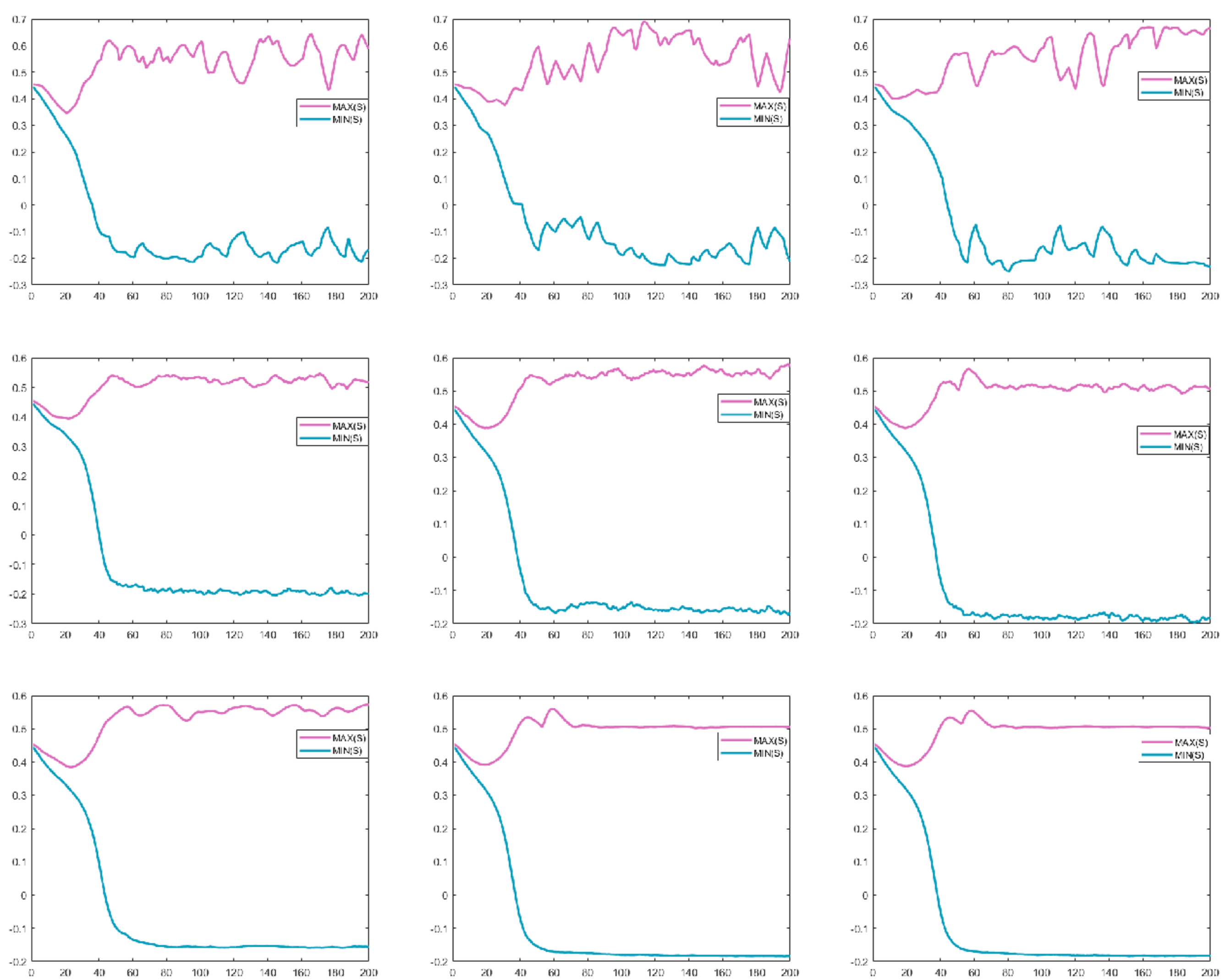

**FIG. 6.** The time evolution of the susceptible individuals with distinct oscillation frequency. The model parameters are consistent with Fig. 4. The transition frequencies of the suspicious individuals' diffusion environment in the first, second, and third rows are 0.01, 1, and 100, respectively. The transition frequencies of the infected individuals' diffusion environment in the first, second, and third columns are 0.01, 1, and 100, respectively.